# Reconfigurable Knots and Links in Chiral Nematic Colloids

**Uroš Tkalec[1†*], Miha Ravnik[2,3], Simon Čopar[3], Slobodan Žumer[3,1], Igor Muševič[1,3*]**

[1]Condensed Matter Physics Department, Jožef Stefan Institute, Jamova 39, 1000 Ljubljana, Slovenia
[2]Rudolf Peierls Centre for Theoretical Physics, University of Oxford, Oxford OX1 3NP, United Kingdom
[3]Faculty of Mathematics and Physics, University of Ljubljana, Jadranska 19, 1000 Ljubljana, Slovenia
[†]Present address: Max-Planck-Institute for Dynamics and Self-Organization, Am Faßberg 17, 37077 Göttingen, Germany

[*]To whom correspondence should be addressed: uros.tkalec@ijs.si (U.T.); igor.musevic@ijs.si (I.M.)

**Tying knots and linking microscopic loops of polymers, macromolecules, or defect lines in complex materials is a challenging task for material scientists. We demonstrate the knotting of microscopic topological defect lines in chiral nematic liquid crystal colloids into knots and links of arbitrary complexity by using laser tweezers as a micromanipulation tool. All knots and links with up to six crossings, including the Hopf link, the Star of David and the Borromean rings are demonstrated, stabilizing colloidal particles into an unusual soft matter. The knots in chiral nematic colloids are classified by the quantized self-linking number, a direct measure of the geometric, or Berry's, phase. Forming arbitrary microscopic knots and links in chiral nematic colloids is a demonstration of how relevant the topology can be for the material engineering of soft matter.**

Knots are fascinating topological objects and symbols of complexity that have fascinated the human mind since the dawn of our history. Although they are treated within the mathematical discipline of topology (*1*), knots and links have always played a prominent role in physical and life sciences (*2*). In supramolecular chemistry, complex links were materialized as interlinked molecular rings – catenanes – and interlocked molecules – rotaxanes (*3-5*). Knotting and the entanglement of polymer molecules were proved to be essential for the crystallization and rheological properties of polymers (*6*). Knotted structures have also been predicted in classical field theory (*7*), and it was recently demonstrated that the threads of zero intensity in interfering beams of light can be knotted and linked as well (*8, 9*). In biological systems, molecular knots and links are particularly important because the entanglement of DNA molecules plays the crucial role in vital life processes of replication, transcription and recombination (*10, 11*). Knot-like topological defects have been observed in chiral nematic liquid crystal (NLC) (*12*) but have remained unexplored because of the difficulty associated with control. The smallness of the length scales involved and the inherent lack of precise control and means of manipulating the knots and links are the major obstacles in studying the structure, properties and mechanisms of their formation.

We demonstrate knotted and linked microscopic loops of topological defects of arbitrary complexity in chiral nematic colloids. The loops are responsible for the stabilization of colloidal structures in a chiral NLC, thus forming an unusual colloidal soft matter (*13-16*). We performed knot and link manipulation by cutting, fusing, and reversibly reconnecting individual defect loops into knots and links of arbitrary complexity using the highly focused light of laser tweezers, which gives us full control over the knot and link formation.

The medium that supports our knots and links is a NLC with colloidal inclusions, and the strings used to tie knots and links are closed defect loops in the NLC. When spherical particles that promote alignment of NLC molecules normal to the surface are dispersed in the NLC, the direction of molecular alignment – the director – is forced to align normal to the curved and closed surface of each inclusion. Because the spherical surface makes it impossible for the molecules to fill the space uniformly, defects in the form of singular points – and in our case, closed defect loops – are created. Each particle is encircled by its own micro-loop, known as a Saturn's ring, in which the degree of molecular order is reduced in the ~10-nm-thick core, and the director exhibits fast spatial variations, making the rings visible under an optical microscope (*16*). The Saturn's ring behaves as an elastic strip that can be stretched and deformed with laser tweezers (*17-20*). More importantly, several Saturn's rings can be fused together using the laser tweezers to entangle a pair or multiple colloidal particles (*21, 22*). Here, the particles and defect loops are topologically and energetically interlinked because the loops must compensate for the topological charge of the particles (*16*) and tend to be as short as possible in order to reduce the total free energy. Although in nematic colloids with a generally homogeneous director alignment only linear entangled objects were successfully created (*22*), the defect loops in chiral NLC colloids can be optically knotted into knots and links of arbitrary complexity.

A dispersion of 4.72-µm-diameter silica microspheres in a pentylcyanobiphenyl (5CB) NLC is used in all the experiments. The surface of the microspheres is chemically functionalized to induce a strong perpendicular alignment of the NLC molecules. The colloidal dispersion is confined to a thin cell, made of glass plates, coated with transparent indium tin oxide and rubbed polyimide alignment layers, spaced by a 5.5-µm-thick mylar foil (*23*). The alignment directions at the top and bottom of the cell are set perpendicular to ensure a 90° twist of the director, thus creating either left- or right-twisted liquid crystal profiles. Using twisted director structure is essential for the stability of nematic knots and links because the cell-imposed twist energetically favors effectively longer and out-of-plane deformed defect loops, which can more easily interlink.

An individual colloidal particle in the chiral nematic cell acquires a single defect loop, shown in Fig. 1A, known as a Saturn's ring (*16*). The ring is tilted at 45° with respect to the molecular alignment at the cell walls and it is clearly visible between crossed polarizers because its core scatters light. In terms of topology, a closed loop without a knot in it is an unknot (*1, 24*). We used laser tweezers to bring together several particles and observed either spontaneous or laser-assisted fusion of their rings, leading to the formation of longer loops that entangle two or more particles (Fig. S1 and Movies S1 to S3). In small colloidal clusters, presented in Fig. 1, B to E, all the loop conformations are likewise topologically equivalent to the unknot.

The simplest nontrivial topological configuration that is created by a sequence of local, isotropic-to-nematic, temperature, and optically induced micro-quenches, is the Hopf link (Fig. 1F). Two interlinked loops, entangled around four neighboring particles, are visible in both the polarizing optical micrograph and the numerically calculated structure.

However, the true richness of the knots and links is revealed when the colloidal clusters are extended to arrays of $p \times q$ particles (Fig. 1G). The laser-assisted knitting technique was applied at multiple knitting sites so as to connect the neighbouring defect rings. A series of nematic braids, realized on $3 \times q$ particle arrays is shown in Fig. 1, G to J (left). To identify the topology of the entangled loops, we performed a sequence of topology-preserving Reidemeister moves (*1*), which virtually transform the real physical conformation of the loops into its planar projection with the minimum number of crossings. Note that only negative or left-handed crossings (*1*) are favored in a left-twisted nematic profile because of the geometric constraint of the cell. The relaxation mappings, illustrated in Fig. 1, G to J (right), reveal a surprising result. There is a series of alternating torus knots and links (*1*): the trefoil knot, the Solomon link, the pentafoil knot, and the Star of David. This generically knotted series of knots and links shows that the confining lattice of colloidal particles allows for the production of torus links and knots of arbitrary complexity, simply by adding and interweaving additional rows of particles – that is, by increasing $q$.

The knots and links can also be reversibly retied. Topologically, this corresponds to locally changing the mutual contact – the unit tangle (*1*) – between the two segments of the knotted line, which can either cross or bypass one another in two perpendicular directions. We were able to reknot the disclination lines in the region of the selected tangle by applying the laser-induced micro-quench, as shown in Fig. 2, thus transforming the unit tangles one into another and consequently changing the topology of the presented conformations. Starting from a tangle inside the encircled region in Fig. 2A, the laser beam initially cut the tangle, and then by using precise positioning and intensity tuning of the beam, the line segments were reknotted into a distinct tangle (Fig. 2C). Further, we reknotted a tangle (Fig. 2C) into another distinct tangle (Fig. 2E). We can make exactly three tangles by reversibly transforming them, one into another. These local transformations change the topology and the handedness (right, +; left, -) of the chiral knots and links, which in a given example corresponds to conversions between right-handed trefoil knot $3_1^+$ (Fig. 2A), the left-handed composite knot $3_1 \# 3_1^-$ (Fig. 2C), and two-component link $6_3^2$ (Fig. 2E). Eventually, the reknotting of knots and links can be performed for any desired knotting sequence of unit tangles at any specific position in a colloidal array. More specifically, the $p \times q$ array of particles generates a template of $(p-1) \times (q-1)$ unit tangles – for example, six tangles on a $4 \times 3$ particle array, which can all be individually switched, thus inducing site-specific transformations between various knots and links.

The optical retying of knots and links is directly related to the changes in the orientational field of the nematic host. Each tangle has four free ends of two defect line segments (insets in Fig. 2, A, C, and E, insets), which in our system form the corners of an approximate tetrahedron (Fig. 2B). The director field inside the tetrahedron has an intrinsic dihedral symmetry, with two perpendicular mirror symmetry planes and a full tetrahedral symmetry on its surface. Therefore, when enclosing any of the three unit tangles with such tetrahedrons the orientational profiles within the tetrahedrons prove to be equivalent, and the only difference is the relative orientation of the tetrahedrons. Consequently, the rewiring of the tangles corresponds to a solid rotation of the director field inside the tetrahedron (*25*), as schematically illustrated in Fig. 2, B, D, and F. By symmetry, these rotations preserve the continuity of the nematic orientational profile and position the two line segments exactly in one of the three distinct configurations, which were recognized as unit tangles.

The defect loops that we used to knit the knots and links are not only structureless strings but are dressed by the surrounding director field. For this reason, they possess a threefold rotational symmetry of the hyperbolic molecular orientational profile in the plane perpendicular to the defect line (Fig. 3A, inset). This threefold symmetric pattern can twist along the disclination and makes the defect lines the three-sided strips, which are analogous to the well-known Moebius strip (*24*). Because our three-sided strips are always closed into loops, only fractional values of the internal twist are allowed by the continuity of the director field, corresponding to the integer multiples of a $2\pi/3$ rotation. This property can be described by a topological invariant called the self-linking number *SL* that counts the turns of the binormal around the curve tangent for algebraic curves (*26*) and can be generalized to count the turns of the director field around the defect loop (*25*). For multiple loops, the summation of *SL* over all the loops can be used as a natural generalization of this invariant. The quantization of the self-linking number is directly related to the geometric, or Berry's, phase (*27*). For nematic braids, the geometric phase corresponds to the angular shift $\varphi$ that is acquired upon the parallel transport of a surface normal travelling along one surface of the characteristic three-sided strip. More specifically, the angular shift is equivalent to $\varphi = 2\pi SL$ (Fig. 3A). Eventually, the *SL* and the number of loops *N* can be used for a unique classification of all possible loop conformations – that is, all the available knot types on a given $p \times q$ particle array. Figure 3B shows the classification of the topological objects on a $3 \times 4$ particle array, determined by testing all the possible combinations of unit tangles. Using *SL* and *N* as the characteristic invariants, the knots and links arrange hierarchically and regularly alternate between the knotted/unknotted and linked/unlinked structures (*25*), which is promising for predicting the complexity of the knots and links that can be realized on a specific $p \times q$ particle array.

Having experimental control over the knotting and having theoretical tools for finding all possible conformations of nematic braids, we were able to perform a made-to-order assembly of knots and links, illustrated in Fig. 3C. First, we selected Borromean rings (*23*) as an example of a complex interwoven structure, which we attempted to weave. Next, we specified the size of the particle array to be $4 \times 4$, the smallest array required for the chosen link. By using a computer algorithm that is based on calculating the polynomial invariants from knot theory (Fig. S2) (*23, 28, 29*), the tangle configuration that corresponds to the chosen link was identified. The site-specific pattern of tangles shown in Fig. 3C (fourth frame) was needed. Lastly, we experimentally assembled the Borromean rings using laser tweezers. To show the reach of our assembly method, we present all of the prime knots and links that can be made-to-order on a $4 \times 4$ particle array in Fig. 3D. Almost 40 different knot and link types are discerned among $3^{(p-1)(q-1)} = 3^9$ tangle combinations with minimum crossing numbers up to 10. Out of all the loop conformations, 35% are prime knots or links, 29% are unknots, 18% are unlinks, and 18% are more complex composite links (Table S1). Such a large diversity of topological objects suggests that it is possible to design any knot or link on a sufficiently large colloidal array.

We have shown that chiral nematic colloids are stabilized by defect knots and links of fascinating complexity, which can be fully controlled and rewired by light. This unusual colloidal soft matter system provides a robust made-to-order assembly of an arbitrary knot or link on a microscopic scale and is a new route to the fabrication of micro-objects with special topological features. We believe that the strategy presented here offers guidance to further progress in our understanding of the knotting of topologically nontrivial entities, such as DNA (*5*), skyrmion lattices in chiral magnets (*30*) and confined blue phases (*31*), and entangled vortices in superconductors (*32*).

**Acknowledgments**

U.T. thanks S. Herminghaus, S. Kralj and S. Vrtnik for discussions and kindly acknowledges support of the Max Planck Society. M.R. acknowledges support of the European Commission (EC) under the Marie Curie Program Active Liquid Crystal Colloids (ACTOIDS); content reflects only the authors views and not the views of the EC. The research was funded by Slovenian Research Agency under the contracts P1-0099, PR-00182 and J1-9728, and in part by the NAMASTE Center of Excellence.

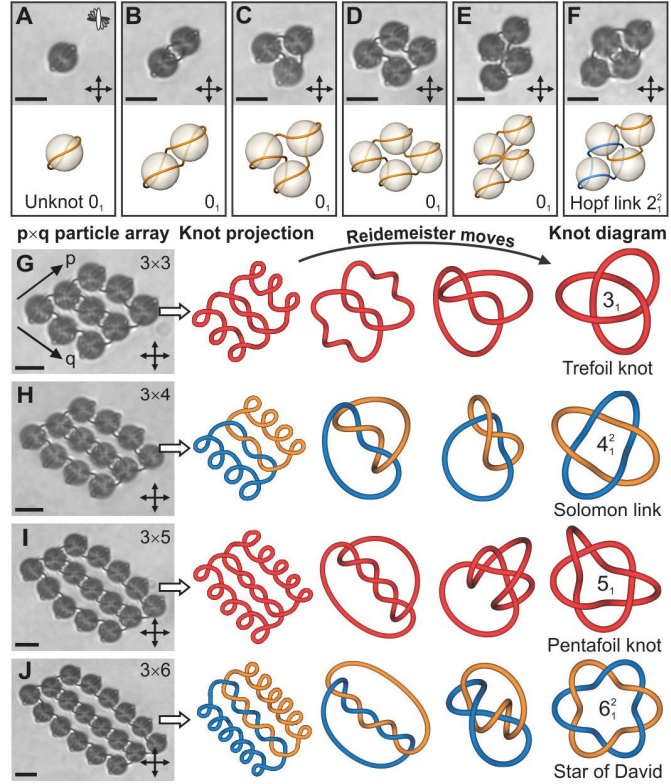

**Fig. 1.** Topological defect lines tie links and knots in chiral nematic colloids. (**A**) A twisted defect ring is topologically equivalent to the unknot and appears spontaneously around a single microsphere. The molecular orientation on the top and bottom of the cell coincides with the orientation of the crossed polarizers. (**B** to **E**) Defect loops of colloidal dimer, trimer and tetramers are equivalent to the unknot. (**F**) The Hopf link is the first nontrivial topological object, knitted from two interlinked defect loops. In (A) to (F), the corresponding loop conformations were calculated numerically by using the Landau–de Gennes free energy model (*13*). (**G** to **J**) A series of alternating torus knots and links on $3 \times q$ particle arrays are knitted by the laser-induced defect fusion. The defect lines are schematically redrawn using a program for representing knots (*33*) to show the relaxation mapping from the initial planar projection to the final knot diagram, which was performed by the sequence of Reidemeister moves. The designations of knots follow the standard notation $C_i^N$, where $C$ indicates the minimal number of crossings, $i$ distinguishes between different knot types, and $N$ counts the number of loops in multi-component links. Scale bars, 5 µm.

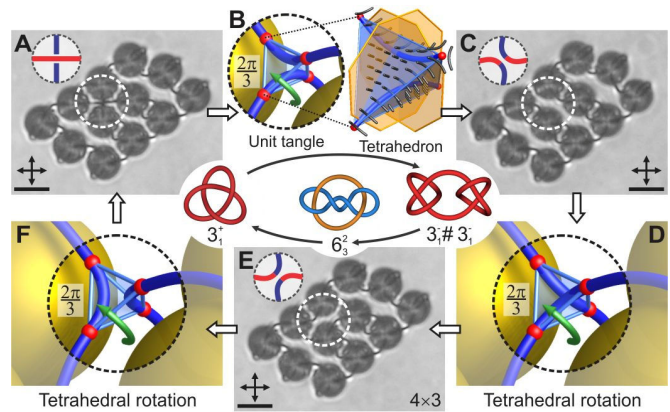

**Fig. 2.** Rewiring of knots and links by use of laser tweezers. (**A**) A right-handed trefoil knot is realized on a $4 \times 3$ colloidal array. The dashed circles indicate a unit tangle that can be rewired with the laser beam. The tangle consists

of two perpendicular line segments and the surrounding molecular field. (**B**) By rewiring the unit tangle that corresponds to a 2π/3 rotation of the encircled tetrahedron, a new composite knot, shown in (**C**), is knitted. The sequence of tangle rewirings in (B), (D), and (F) results in switching between knots and links, demonstrated in (A), (C), and (E). Scale bars, 5 μm.

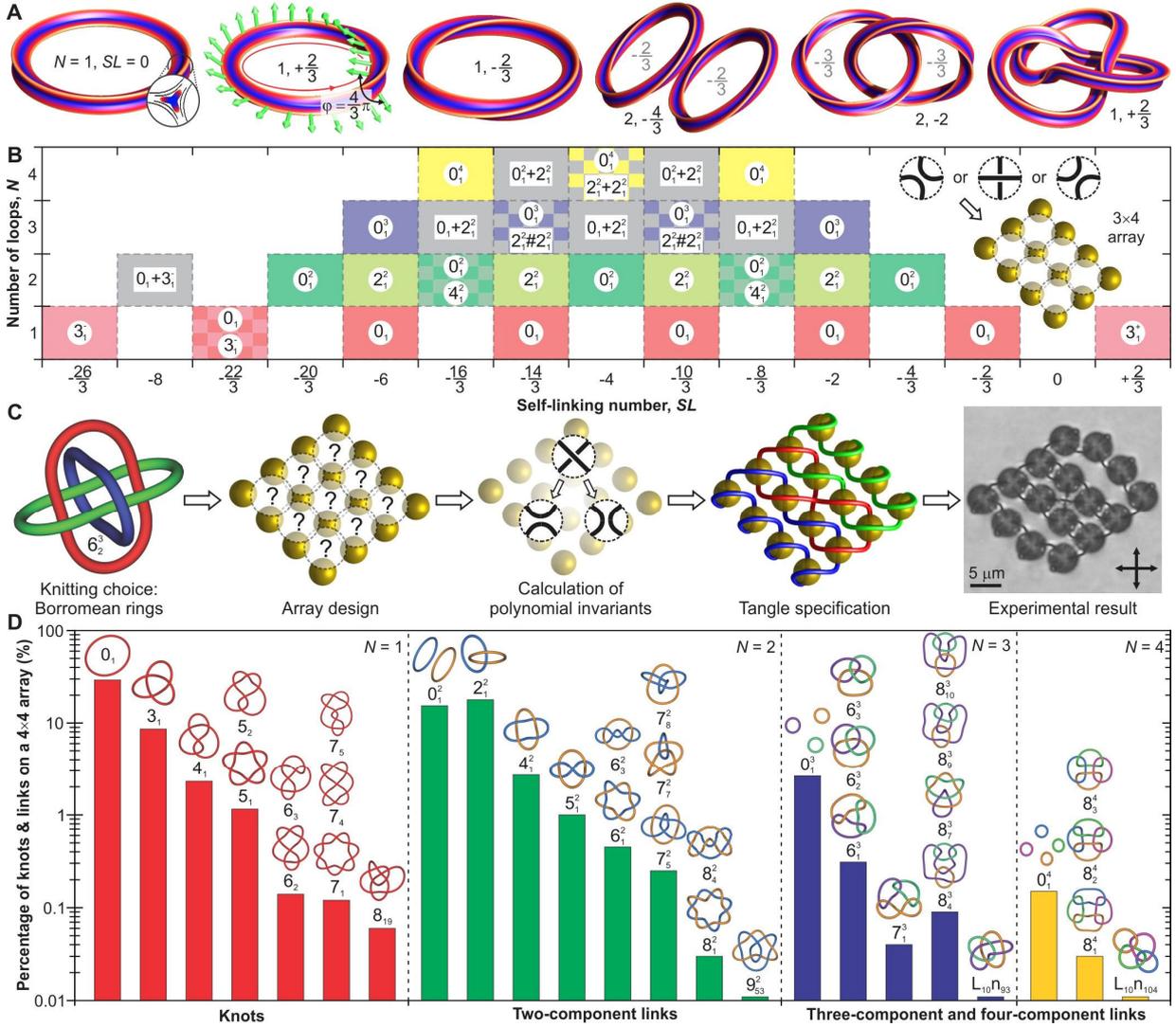

**Fig. 3.** Topological classification and made-to-order assembly of linked and knotted nematic braids. (**A**) The defect loops have a local threefold rotational symmetry of the hyperbolic cross-section and correspond to three-sided strips, analogous to the Moebius strip. They can be distinctly characterized by the number of loops *N* and the self-linking number *SL* of fractional values. (**B**) The classification of all possible knots and links on a 3 × 4 particle array by *SL* and *N*. The hierarchical ordering of knots and links, depicted by distinct colors and standard knot symbols, are shown. (**C**) Made-to-order assembly of Borromean rings on a particular $p \times q$ particle array. The feasible tangle combinations were tested by the numerical algorithm based on the Jones polynomials and Kauffman bracket approach (*24*, *28*). The selected configuration was identified by direct comparison with polynomials in the enumerated *Table of Knot Invariants* (*29*) and then assembled by using laser tweezers. (**D**) The distribution of prime knots and links on a 4 × 4 particle array shows a large diversity of topological conformations with minimum crossing numbers up to 10. The probability of occurence of a particular knot or link decreases with its complexity, as measured by the minimum crossing number.